\documentclass[aps,prl,amsmath,amssymb,amsfonts,nofootinbib,long]{revtex4}
\usepackage{graphicx}

\newcommand\M{\mbox{\scriptsize max}}
\newcommand{\Pl}{\mbox{\scriptsize Pl}}

\newcommand{\br}{\langle}
\newcommand{\kt}{\rangle}
\newcommand\G{\mbox{G}}

\newcommand\eV{\mbox{eV}}

\newcommand\MeV{\mbox{MeV}}
\newcommand\GeV{\mbox{GeV}}

\newcommand\aem{\alpha_{\rm em}}

\newcommand\U{U(1)_{\rm PQ}}
\newcommand\f{f_{12}}

\newcommand\T{\Theta}

\begin{document}

\title{A Note on the Cosmic Evolution of the Axion in a Strong Magnetic Field}

\author{L. Campanelli$^{1,2}$}
\email{campanelli@fe.infn.it}
\author{M. Giannotti$^{1,2}$}
\email{giannotti@fe.infn.it}

\affiliation{$^{1}${\it Dipartimento di Fisica, Universit\`a di Ferrara, I-44100 Ferrara, Italy
\\           $^{2}$INFN - Sezione di Ferrara, I-44100 Ferrara, Italy}}

\date{August 2006}

\begin{abstract}
It has been pointed out in the literature that in the presence of
an external magnetic field the axion mass receives an
electromagnetic contribution. We show that if a magnetic field
with energy density larger than $\sim 10^{-8}$ times the energy
density of the Universe existed at temperatures of a few $\GeV$,
that contribution would be dominant and consequently the cosmic
evolution of the axion field would change substantially. In
particular, the expected axion relic abundance would be lowered,
allowing a small relaxation of the present cosmological bound on
the Peccei-Quinn constant.
\end{abstract}

%*************************************************************************************************%

\maketitle

\vspace*{0.3cm}

{\bf Keywords:} Axions, Magnetic Fields

\vspace*{1.5cm}

%***************************************   Body   ************************************************%

The existence of the axion field is predicted by the Peccei-Quinn
(PQ) mechanism \cite{Pec77} for the solution of the strong CP
problem, one of the most puzzling points of modern particle
physics (for a review see, e.g., \cite{Kim87}). This problem is
related to the presence of the P- and CP-violating term
$\mathcal{L}_{\T}=(\alpha_s/8 \pi) \T \, G\Tilde G$, known as
$\T-$term, in the QCD Lagrangian. Here, $\alpha_s$ is the fine
structure constant of the strong interactions, while $G$ and
$\Tilde G$ are the gluon field and its dual. In the PQ-mechanism
the parameter $\T$ becomes a dynamical field, the axion itself
$a=\T f_a$, which emerges as the (pseudo-)Goldstone mode of the
PQ-symmetry $\U$, spontaneously broken at the energy scale $f_a$.
The parameter $f_a$, known as the PQ- or axion-constant,
characterizes the all axion phenomenology \cite{Sik05}, and is
presently constrained in the very narrow region $10^9 \lesssim f_a
\lesssim 10^{12} \GeV$ by astrophysical and cosmological
considerations \cite{Raff}. The axion potential, generated by the
non trivial axion-gluon interactions, is minimized for the CP-even
configuration $\br a\kt=0$, providing therefore a dynamical
explanation for the CP-conserving behavior of the strong
interactions.

The cosmic evolution of the axion field is described by the
equation \cite{Kolb}
\begin{equation}\label{1}
\ddot{\Theta} + 3H \dot{\Theta} + m_a^2(T) \Theta = 0,
\end{equation}
where $H \simeq 1.66 g_{*}^{1/2} T^2/m_{\Pl}$ is the Hubble
parameter (in the radiation era) with $g_{*}$ the total number of
effectively massless degrees of freedom and $m_{\Pl}$ the Planck
mass. The temperature-dependent axion mass is \cite{GPY} (see also
\cite{Kolb})
\begin{equation}
\label{m} m_a(T) \simeq
       \left\{
        \begin{array}{ll}
         0.1 m \left(\Lambda/T\right)^{3.7} & (T\gg \Lambda), \\
             m                              & (T \ll \Lambda),
        \end{array}
       \right.
\end{equation}
where $m$ is the zero temperature limit
$m \simeq 6.2 \times 10^{-6} \eV / \f$, $\f = f_{a}/(10^{12}
\GeV)$,
and $\Lambda \sim 200 \MeV$ the QCD scale.

For temperatures high enough for the mass term in Eq.~(\ref{1}) to
be negligible, the axion has no dynamics, and therefore $\T$
remains fixed on its initial value $\T_i$, which is not required
to be zero. However, as the axion mass becomes dominant on the
friction (Hubble) term, $\Theta$ begins to oscillate with the
frequency $m_a(T)$ and will eventually approach the CP-conserving
limit $\T_{\rm today} \sim 0$. During this period of coherent
oscillations the number of axions in a comoving volume remains
constant and therefore the axion relic abundance today can be
straightforwardly evaluated as
\begin{equation}
\label{Omega} \Omega_a \simeq 1.6 \, \T^2_i \, g_{*1}^{-1/2}
f_{12} \, \frac{\GeV}{T_1} \, ,
\end{equation}
where the temperature $T_1$, defined by the equation $m_a(T_1) =
3H(T_1)$, indicates approximately the time when the oscillations
start, and $g_{*1} = g_{*}(T_1)$.

If the only contribution to the axion mass were given by the QCD
instanton effects (\ref{m}), then $T_1 \simeq 0.9 \,
\Lambda_{200}^{0.65} f_{12}^{-0.175} \, \GeV$, where
$\Lambda_{200} = \Lambda/200\MeV$. In this approximation
Eq.~(\ref{Omega}) reduces to
$\Omega_{a} \simeq 0.2 \, \Lambda_{200}^{-0.65} \Theta_i^2
\f^{1.175}$.
With the natural choice $\Theta_i \simeq 1$ \cite{Kolb}, this
gives $\Omega_{a} \simeq 0.3$ (the expected dark matter abundance)
for $\f \simeq 1$. Much larger values of $\f$ would cause too much
axion production and are therefore excluded. This observation
leads to the upper limit on the PQ-constant mentioned above, the
so called cosmological bound $f_a \lesssim 10^{12} \GeV$
\cite{Pre83}.

However, as pointed out in Ref.~\cite{VMP}, in an external uniform
magnetic field $B \gg B_c \simeq 4.4 \times 10^{13} \G$ ($B_c$ is
the critical or Schwinger value) the axion mass receives an
electromagnetic contribution,
\footnote{A uniform magnetic field, besides giving a contribution
to the axion mass, causes a dissipation of the axion field itself.
This is induced by the axion-photon conversion in the magnetic
field. However, this affects only negligibly the expected axion
relic abundance \cite{Ahon96}.}
\begin{equation}
\label{deltamB}
\delta m_a(B) \simeq 5.8 \, \xi \left(
\frac{B}{10^{23} \mbox{G}} \right )^{\! 1/2} \: \frac{10^6
\GeV}{f_a} \: \eV,
\end{equation}
where $\xi$ is a model-dependent parameter of order of unity
related to the effective axion photon coupling $g_{a\gamma} = \xi
\aem /(2\pi f_a)$.
(It is worth noting that the total axion mass is given by $m_{\rm
tot}^2 = m_a^2 + \delta m_a^2$.)
As we will show below this result has important consequences for
axion cosmology. Indeed, if a magnetic field with energy density
larger than about $10^{-8}$ times the energy density of the
Universe existed before the period of the axion coherent
oscillations, the electromagnetic contribution to the axion mass
would be dominant, and the beginning of the oscillations would
consequently start earlier. The most important consequence of this
result is a reduction of the expected axion relic abundance [see
Eq.~(\ref{Omega})], and therefore a relaxation of the cosmological
upper bound on the PQ-constant.

The existence of very intense magnetic fields in the early
Universe is not excluded \cite{Magn1,Magn2}. Indeed, it has been
invoked for the explanation of the presently observed large-scale
magnetic fields, and could have interesting repercussions on
the axion phenomenology (see, e.g., \cite{Csaki}).
Since the primordial plasma is an excellent conductor, magnetic
fields are frozen into the plasma and evolve as $B \propto T^2$.
Introducing the constant $b$ as $B = b T^2$, we can parameterize
the evolution of the magnetic field as $B \simeq 1.4 \times
10^{19} b \, (T/\GeV)^2 \, \mbox{G}$. Requiring the magnetic
energy density $\rho_B = B^2/2$ to be less than the energy density
of the Universe in the radiation era $\rho = (\pi^2/30) g_{*} \,
T^4$ (to be consistent with the constraint on primordial magnetic
fields coming from the Big Bang Nucleosynthesis and the Cosmic
Microwave Background \cite{Magn2}), we find the maximum allowed
value of $b$, $b_{\M} \simeq 0.8 \, g_{*}^{1/2}$.

It is useful to re-write Eq.~(\ref{deltamB}) as
\begin{equation}
\label{deltamT} \delta m_a(T) \simeq 0.2 \times 10^{-2} \xi \,
b^{1/2} \Lambda_{200} \, m \, \frac{T}{\Lambda} \, ,
\end{equation}
and to introduce the temperature $T_*$ such that the QCD axion
mass (\ref{m}) and the electromagnetic contribution
(\ref{deltamT}) are equal, $\delta m_a(T_*) = m_a(T_*)$. It
results in $T_* \simeq 2.2 \, \xi^{-0.2} b^{-0.1}
\Lambda_{200}^{-0.2} \Lambda$. Let us continue to represent the
temperature at which the axion field begins to oscillate as $T_1$.
\\
For $b \lesssim b_* = 0.9 \times 10^{-3} \, \xi^{-2}
\Lambda_{200}^{1.5} \, \f^{1.75}$ we have $T_1 < T_*$, thus the
electromagnetic contribution to the axion mass is negligible with
respect to that of QCD [see Eqs.~(\ref{m}) and (\ref{deltamT})].
Therefore the standard analysis applies.
\\
However, if $b \gtrsim b_*$ (which for $\Lambda_{200} = \xi = \f
=1$ corresponds to a magnetic energy $\rho_B \gtrsim 10^{-8}
\rho$) the electromagnetic contribution dominates and then $T_1$
is determined by imposing that $\delta m_a(T_1) = 3H(T_1)$. In
this case we find $T_1 \simeq 0.2 \times 10^3 \, \xi \, b^{1/2}
f_{12}^{-1} g_{*1}^{-1/2} \, \GeV$. Then, inserting the value of
$T_1$ in Eq.~(\ref{Omega}) we get
\begin{equation}
\Omega_a \simeq 0.9 \times 10^{-2} \Theta^2_i \, \xi^{-1} b^{-1/2}
f_{12}^{2} \, .
\end{equation}

Requiring $\Omega_a \lesssim 0.3$, we get that the maximum value
of the Peccei-Quinn constant is $f_{12} \simeq 5.8 \, \xi^{1/2}
b^{1/4} \Theta^{-1}_i$ to which corresponds the temperature $T_1
\simeq 29.6 \, \xi^{1/2} b^{1/4} \Theta_i \, g_{*1}^{-1/2} \,
\GeV$. Because $\xi$ and $\Theta_i$ are of order of unity, taking
$b = b_{\M}$ we get $f_{12} \simeq 9.6$ and $T_1 \simeq 5.3 \,
\GeV$.
%[Note that for $\Lambda_{200} = \xi = 1$, we have $T_* \simeq 0.4
%\, \GeV$, and $m_a(T_1) / \delta m_a(T_1) \simeq 3 \times
%10^{-6}$.]

In conclusion, we have shown that a strong cosmological magnetic
field can have a non-negligible influence on axion cosmology. In
particular, the cosmological limit on the axion constant could be
relaxed by one order of magnitude. In this case, the axion
interactions with matter and photons would be reduced, rendering
the axion more ``invisible''. As a final observation we note that
the electromagnetic contribution to the axion mass,
Eq.~(\ref{deltamB}), was computed in the zero temperature limit
and so our conclusions must be considered only as a preliminary
result. Indeed, a careful calculation of the mass shift at finite
temperature is in progress.
Whatever the case, the phenomenon of axion mass shift in a strong
magnetic field discussed in Ref.~\cite{VMP} needs to be considered
seriously since it seems to be the most relevant effects of a
uniform magnetic field on axion cosmology.

%************************************   Acknowledgments   ****************************************%

\vspace*{0.5cm}

We would like to thank Elizabeth T. Price for carefully reading
the manuscript.

%*************************************   Bibliography   ******************************************%

\end{document}